\documentclass[twocolumn,secnumarabic,amssymb, nobibnotes, aps, prd]{revtex4}
\usepackage{graphicx}

\begin{document}
\title{Reply to Comment on 'The Law of Entropy Increase and the Meissner Effect'}
\author{A. V. Nikulov}
\affiliation{Institute of Microelectronics Technology and High Purity Materials, Russian Academy of Sciences, 142432 Chernogolovka, Moscow District, RUSSIA} 

\begin{abstract} Jorge Hirsch, in his Comment on my article published in Entropy, agrees that the conventional BCS theory of superconductivity contradicts the second law of thermodynamics. He tries to prove that this theory cannot be valid because of this contradiction, since the Meissner effect is consistent with the second law of thermodynamics, according to his alternative theory of superconductivity, which can explain how the persistent currents can be annihilated before the resistance appearance. First of all I draw attention on experimental facts according to which the persistent currents observed in rings can be not annihilated not only before but even after the resistance appearance, in accordance with the conventional theory of superconductivity and contrary to the second law of thermodynamics. Hirsch's proof of the consistency of the Meissner effect with the second law of thermodynamics is based on mistakes provoked by ignoring the contradictions between books on superconductivity and the false analogy of the Gorter cycle with the Carnot cycle. In fact, Hirsch did not refute, but confirmed the conclusion made in my article about the contradiction of the Meissner effect to the second law of thermodynamics.

 \end{abstract}

\maketitle


\section{Introduction}
Jorge Hirsch argues in his Comment \cite{Hirsch2024Comment} on my article \cite{Entropy2022} that the conventional BCS theory of superconductivity contradicts to the second law of thermodynamics. I agree with Hirsch in this. Moreover, I noticed this contradiction thanks to Hirsch's publications \cite{HirschEPL,HirschIJMP,HirschPhys}. I was sure that the persistent currents are damped with the generation of Joule heat because of a non-zero resistance in the normal state. I knew also that the persistent currents can reappear when returning to the superconducting state. But I, like all experts on superconductivity, kind of forgot that the the generation of Joule heat is an irreversible process according to the second law of thermodynamics. Thanks to Hirsch's publications, I learned that ninety years ago physicists knew this basis of thermodynamics. Before the discovery of the Meissner effect in 1933 \cite{Meissner1933} "{\it it was assumed that the transition in a magnetic field is substantially irreversible, since the superconductor was considered to be a perfect conductor, in which, when the superconductivity is destroyed, the surface currents associated with the field are damped with the generation of Joule heat}" \cite{Shoenberg1938}. But since, according to this notion, the Meissner effect refutes the second law of thermodynamics, all physicists after 1933 are sure that this transition is the first-order phase transition.

\section{The persistent currents may not be annihilated not only before, but also after the material gets resistance}
Some physicists knew ninety years ago that "{\it the conception of Joule-heat can rather difficult be reconciled with reversibility}" \cite{Keesom1934} and therefore "{\it it is essential that the persistent currents have been annihilated before the material gets resistance, so that no Joule heat is developed}" \cite{Keesom1934}. The famous physicist W.H. Keesom actually set in 1934 \cite{Keesom1934} the task for the future theory of superconductivity to explain how the persistent currents can be annihilated before the material gets resistance in order to avoid a contradiction of the Meissner effect with the second law of thermodynamics. The creators of the conventional theory of superconductivity \cite{BCS1957,GL1950} did not even consider the Keesom task \cite{Keesom1934}. They, like most other superconductivity experts, kind of forgot that Joule heating is an irreversible thermodynamic process. 

I agree with Hirsch that only one of the two alternatives can be possible: 1) the second law of thermodynamics is invalid; 2) the conventional theory of superconductivity \cite{BCS1957,GL1950} is invalid. The persistent currents reappears at the return to the superconducting state not only at the Meissner effect, but also at the observations of the flux quantization \cite{fluxquan1961a,fluxquan1961b} and of the persistent current in superconducting rings with the weak screening $s \ll \lambda _{L}^{2}$ \cite{PCJETP07}. The latter is especially important for Hirsch alternatives, since according to experimental results \cite{Moler2007,Letters2007} obtained for the first time in 1962 \cite{LP1962}, in the case of the weak screening, the persistent current may not be annihilated not only before, but also after the material gets resistance. The article \cite{PhysicaC2021} emphasizes that even solving the Keesom task cannot save the faith in the second law of thermodynamics due to the observation of the persistent current \cite{Moler2007,Letters2007,LP1962} that is not damped at thermodynamic equilibrium despite a nonzero resistance $R > 0$ and a nonzero power of dissipation $RI_{p}^{2}$. 

This paradoxical phenomenon, observed in a narrow fluctuation region near the temperature of superconducting transition $T \approx T_{c}$, was explained \cite{PRB2001} in the framework of the conventional theory of superconductivity \cite{BCS1957,GL1950} as a directed Brownian motion. More than a hundred years ago, physicists knew that the impossibility of directed Brownian motion \cite{Smoluchowski} or the assumption of molecular disorder \cite{Planck} is a condition for the validity of the second law of thermodynamics. The directed Brownian motion observed in a nonuniform superconducting ring at $T \approx T_{c}$ where $R > 0$ \cite{Moler2007,Letters2007,LP1962} challenges the second law of thermodynamics since the persistent current $I_{p} \neq 0$ can create a dc voltage $V_{p} \propto I_{p}$ and a dc power $V_{p}I_{p}$ on segments of this nonuniform ring \cite{LTP1998,Berger2024PRB} according to the conventional theory of superconductivity \cite{BCS1957,GL1950}. The dc power $RI_{p}^{2}$ \cite{Moler2007,Letters2007,LP1962} and $V_{p}I_{p}$ \cite{PLA2012Ex} observed at thermodynamic equilibrium on superconducting rings at $T \approx T_{c}$ where $R > 0$ is experimental evidence against the second law of thermodynamics and in favor of the conventional theory of superconductivity \cite{BCS1957,GL1950}. 

According to the conventional theory of superconductivity \cite{BCS1957,GL1950} the persistent currents are damped because of a non-zero resistivity and reappear because of quantization both in a ring with the weak screening and in a bulk superconductor, i.e. the macroscopic persistent currents must appear in all cases when superconducting state without currents is forbidden because of quantization \cite{PhysicaC2021}. Thus, according to the conventional theory of superconductivity \cite{BCS1957,GL1950}, the reason for the violation of the second law of thermodynamics in macroscopic quantum phenomena observed in superconductors is quantization. Hirsch in his Comment \cite{Hirsch2024Comment} disputes the existence of this violation only in the case of the Meissner effect \cite{Meissner1933}. I pay attention to the mistakes and contradictions in his arguments.           

\section{Contradictions between books on superconductivity} 
Hirsch uses in his Comment \cite{Hirsch2024Comment} opposite statements since he ignores the contradictions between books on superconductivity. According to the most books \cite{Shoenberg1938,Shoenberg1952,Kittel956,Lynton1962,Buckel1972,Schmid1997} the free energy is not changes 
$$F_{nH} = F_{sH}  \eqno{(1)}$$ 
at the superconducting state in the critical magnetic field $H = H_{c}(T)$, increases with magnetic field 
$$F_{sH} = F_{s0} + E_{m} = F_{s0} + \frac{V\mu_{0}H^{2}}{2} \eqno{(2)}$$ 
in the superconducting state and does not change 
$$F_{nH} = F_{n0} \eqno{(3)}$$
in the normal state. In opposite to these statements authors of fewer books \cite{Ginzburg1946,Gennes1966,Tinkham1996} state, that the free energy changes 
$$F_{nH} = F_{sH} + A_{sn} \eqno{(4)}$$ 
by the amount of work $A_{sn}$ performed during the transition of a bulk superconductor with a macroscopic volume $V$ to the normal state at $H = H_{c}(T)$, magnetic field does not change the free energy 
$$F_{sH} = F_{s0} \eqno{(5)}$$
of the superconducting state and increases 
$$F_{nH} = F_{n0} + E_{m} = F_{n0} + \frac{V\mu_{0}H^{2}}{2} \eqno{(6)}$$
the free energy of the normal state. V.L. Ginzburg calculated \cite{Ginzburg1946} the negative work $A_{ns} = -2E_{m} =  -V\mu_{0}H_{c}^{2}$ performed at the transition from normal to superconductor state at $H = H_{c}(T)$. This work decreases the total energy of the superconductor and can be used in a load. Other Nobel prize winner, P.G. de Gennes calculated the positive work $A_{sn} = 2E_{m} =  V\mu_{0}H_{c}^{2}$ performed at the opposite transition \cite{Gennes1966}. This work increases the total energy of the superconductor and is performed by the power source of the solenoid. 

Hirsch follows to the most books \cite{Shoenberg1938,Shoenberg1952,Kittel956,Lynton1962,Buckel1972,Schmid1997}, when he assumes that the sample should release the latent heat $L$ at $H = H_{c}(T)$, see the equation (1) in \cite{Hirsch2024Comment}. The latent heat $L = T\Delta S$, equal the product of temperature $T$ by the entropy jump $\Delta S$, can be observed only at the first-order phase transition at which the free energy cannot change since the entropy is the derivative $S = -dF(T)/dT$ of free energy $F$ with respect to temperature $T$. The jump in the derivative $S = -dF(T)/dT$ can be finite only if the function $F(T)$ does not change by a jump. Therefore, the latent heat has a sense only according to the equation (1), which is written in the most books \cite{Shoenberg1938,Shoenberg1952,Kittel956,Lynton1962,Buckel1972,Schmid1997}, but not according to the equation (4), which is written in the fewer books \cite{Ginzburg1946,Gennes1966,Tinkham1996}. According to the equation (4) the free energy $F(T)$ changes by a jump at $T = T_{c}(H)$: $F(T) = F_{sH}$ at $T < T_{c}(H)$ and $F(T) = F_{nH} = F_{sH} + A_{sn} = F_{sH} + V\mu_{0}H_{c}^{2}$ at $T > T_{c}(H)$. 

Since the latent heat does not make sense according to equation (4), Hirsch contradicts himself when he agrees with V.L. Ginzburg \cite{Ginzburg1946} that the negative work $A_{ns} = -2E_{m} =  -V\mu_{0}H_{c}^{2}$ is performed at the transition to superconductor state at $H = H_{c}(T)$, see the equation (9) in \cite{Hirsch2024Comment}. The negative work, decreasing $\Delta U = A_{ns} = - V\mu_{0}H_{c}^{2}$ the total energy $U = F + ST$, must decrease either the free energy $\Delta F = - V\mu_{0}H_{c}^{2}$, the thermal energy $\Delta Q = \Delta ST = - V\mu_{0}H_{c}^{2}$ or the both $\Delta F +  \Delta ST = - V\mu_{0}H_{c}^{2}$. V.L. Ginzburg \cite{Ginzburg1946} and the authors \cite{Gennes1966,Tinkham1996} were sure that the positive $A_{sn} = V\mu_{0}H_{c}^{2}$ and negative $A_{ns} = -2E_{m} =  -V\mu_{0}H_{c}^{2}$ work can change only the free energy (4) since they knew that the decrease of the thermal energy $\Delta ST = - V\mu_{0}H_{c}^{2}$ or $\Delta F +  \Delta ST = - V\mu_{0}H_{c}^{2}$ at the constant temperature $T_{1} = T_{c}(H)$ contradicts to the second law of thermodynamics.     

\section{The falsity of the analogy with the Carnot engine}
Hirsch, unlike the authors \cite{Ginzburg1946,Gennes1966,Tinkham1996}, does not understand that his equation (10) in \cite{Hirsch2024Comment} contradicts to the second law of thermodynamics. Hirsch writes correctly that "{\it the total energy delivered to the power source by the sample of unit volume becoming superconducting is}" $W = \mu_{0}H_{c}^{2}$, see the equation (9) in \cite{Hirsch2024Comment}. The half of this energy is the energy of the magnetic field which decreases from $E_{m} =  V\mu_{0}H_{c}^{2}/2$ to $E_{m} \approx 0$ at the transition to the superconducting state. Only this half decreases the free energy $\Delta F = F_{sH} - F_{nH} = - V\mu_{0}H_{c}^{2}/2$, while the second half (the surplus work \cite{Keesom1934}) $A_{surp} = A_{ns} + E_{m} = -V\mu_{0}H_{c}^{2}/2 = \Delta ST$ is taken from heat, according to the equation (10) in the Hirsch Comment \cite{Hirsch2024Comment}.    

Hirsch does not understand what the authors \cite{Ginzburg1946,Gennes1966,Tinkham1996} understood because of his false analogy with a Carnot engine.
He knows that the total entropy of the universe does not change in the Carnot cycle. But the total entropy does not change $\Delta S_{tot} = \Delta S_{de} + \Delta S_{in} = 0$ due to the irreversible process when the heat $Q_{co}$ flows from the heater with a higher temperature $T_{he}$ to the cooler with a lower temperature $T_{co} < T_{he}$. The increase in the entropy $\Delta S_{in} = Q_{co}/T_{co} - Q_{co}/T_{he} = (Q_{he} - W)/T_{co} - (Q_{he} - W)/T_{he}$ at this irreversible process compensates the decrease in the entropy $\Delta S_{de} = -W/T_{he} =  -(Q_{he} - Q_{co})/T_{he}$ when a part of the heat is used for an useful work $W =  Q_{he} - Q_{co}$ at the heater temperature $T_{he}$. The efficiency $\eta = W/Q_{he}$ of conversion of heat $Q_{he}$ into work $W$ of the Carnot engine
$$\eta_{max} = (\frac{W}{Q_{he}})_{max} = 1 - \frac{T_{co}}{T_{he}}  \eqno{(7)}$$
is deduced from the condition of invariability of the total entropy: $\Delta S_{tot} = \Delta S_{de} + \Delta S_{in} = (Q_{he}/T_{co})(1 - T_{co}/T_{he} - W/Q_{he}) = 0$. 

According to the Carnot principle, which "{\it we call the second law of thermodynamics since Clausius's time}" \cite{Smoluchowski}, the efficiency of any heat engine cannot exceeds the efficiency of the Carnot engine (7). It is obvious that the Carnot principle cannot be violated in the Carnot cycle. Hirsch uses the analogy of the Gorter cycle with the Carnot cycle, but he proves himself that this analogy is false. He writes: "{\it We assume that all the heat exchanges with the environment happen under conditions where the system and the environment are at the same temperature, so that no net entropy is generated by the heat transfer process}" \cite{Hirsch2024Comment}. This assumption $T_{sys} = T_{env}$ means that the decrease in the entropy, which must be according to the equation (10) in \cite{Hirsch2024Comment}, cannot be compensated in the Gorter cycle by the increase in the entropy due to heat flows from a hot body to a cold one:  $\Delta S_{in} = Q/T_{sys} - Q/T_{env} = 0$ at $T_{sys} = T_{env}$. Thus, the Comment \cite{Hirsch2024Comment} is a confirmation rather than a refutation of the conclusion made in the article \cite{Entropy2022} that the Meissner effect refutes the second law of thermodynamics.

\section{Physicists were forced to forget the basics of thermodynamics because of their belief in thermodynamics}
Hirsch's mistakes, as well as the contradictions between books on superconductivity, were provoked by the belief of all physicists in thermodynamics. The belief that the phenomenon of superconductivity cannot contradict thermodynamics appeared a hundred years ago \cite{Lynton1962}. According to this belief, the transition to a superconducting state occurs when the free energy of the superconducting state becomes less than the normal state, $F_{s0} < F_{n0}$ at $T < T_{c}$; $F_{sH} < F_{nH}$ at $H < H_{c}(T)$, and the free energy should be equal at the superconducting transition $F_{s0} = F_{n0}$ at $T = T_{c}$; $F_{sH} = F_{nH}$ at $H = H_{c}(T)$. Equations (4-6), first derived by V.L. Ginzburg in 1946 \cite{Ginzburg1946}, correspond to  thermodynamics and the law of energy conservation but contradict the common belief: the free energies are not equal at $H = H_{c}(T)$ according to (4) and the equality of free energies (1) at $H = H_{c}(T)$ cannot be derived from inequality $F_{s0} < F_{n0}$ at $H = 0$ and equations (5), (6). 

For this reason, the authors of most books \cite{Shoenberg1938,Shoenberg1952,Kittel956,Lynton1962,Buckel1972,Schmid1997} followed equations (1-3), first derived in 1934 by C.J. Gorter and H. Casimer \cite{Gorter1934} rather than equation (4-6). It is surprising that no one noticed for ninety years that in order to derive equations (2), (3) C.J. Gorter and H. Casimer had to use a false claim that "{\it the work, done by the current in the coil, which brings about $H$}" \cite{Gorter1934} creates the energy $dA = -VHdM$ of magnetization $M = B - \mu_{0}H$ rather than the energy of magnetic field $dA = VHdB$. The falsity of this claim is so obvious that it is surprising that this claim could have been done by physicists. According to this claim and contrary to the law of energy conservation, no work is needed in order to create a magnetic field $H$ in the volume $V$ of the empty coil, inside which the magnetization $M = B - \mu_{0}H = 0$. 

The authors of fewer books \cite{Ginzburg1946,Gennes1966,Tinkham1996} understood that the equality (1) cannot be deduced without the contradiction with the second law of thermodynamics because of the work performed at $H = H_{c}(T)$. Replacing the correct expression $dA = VHdB$ with the false one $dA = -VHdM$ by C.J. Gorter and H. Casimer \cite{Gorter1934} changed the sign of the work $A_{sn}$, but could not make the work zero: this work is positive $A_{sn} = V\mu_{0}H_{c}^{2}$ according to P.G. de Gennes \cite{Gennes1966} and is negative $A_{snGC} = - V\mu_{0}H_{c}^{2}$ according to \cite{Gorter1934}. The negative work $A_{snGC} = - V\mu_{0}H_{c}^{2}$ is done by reducing thermal energy according to the equality (11) of the article \cite{Gorter1934}. Thus, "{\it the thermodynamics of superconductors developed by Gorter and Casimir}" \cite{Lynton1962} contradicts not only the law of energy conservation, but also the second law of thermodynamics. Almost all experts on superconductivity believed in this thermodynamics for ninety years because of their belief that superconductivity phenomena cannot contradicts thermodynamics.  

\section{Conclusion}
Jorge Hirsch was the first who drew attention that superconductivity experts had forgotten the basics of thermodynamics \cite{HirschEPL,HirschIJMP,HirschPhys}. But he did not take into account that they forgot not only that the generation of Joule heat is an irreversible thermodynamic process, but also that no work can be performed during a phase transition. The authors of most books \cite{Shoenberg1938,Shoenberg1952,Kittel956,Lynton1962,Buckel1972,Schmid1997} had to contradict the second law of thermodynamics, and the authors of fewer books \cite{Ginzburg1946,Gennes1966,Tinkham1996} had to forget that free energy cannot change at a phase transition since negative work $A_{ns} =  -V\mu_{0}H_{c}^{2}$ should decrease either thermal energy $\Delta ST = - V\mu_{0}H_{c}^{2}$ or free energy $\Delta F = F_{sH} - F_{nH} = - V\mu_{0}H_{c}^{2}$. 

Hirsch, who also forgot this basis of thermodynamics, misunderstood numerous experiments by W. H. Keesom and coworkers, Ref. [6] in the Comment \cite{Hirsch2024Comment}. W.H. Keesom wrote in 1934: "{\it Till now we imagined that the surplus work served to deliver the Joule-heat developed by the persistent currents the metal getting resistance while passing to the non-supraconductive condition. As, however, the conception of Joule-heat can rather difficult be reconciled with reversibility we think now that there must be going on another process that absorbs energy}" \cite{Keesom1934}. W. H. Keesom changed his opinion about the surplus work $A_{surp} = A_{sn} - E_{m} = V\mu_{0}H_{c}^{2}/2$ not on the basis of his measurements, which did not change after 1933, but because he began to consider as reversible the superconducting transition at $H = H_{c}(T)$. Not results of measurements but their interpretation was changed after the discovery of the Meissner effect \cite{Meissner1933} in 1933: the heat was interpreted as the Joule heat before 1933 and as the latent heat after 1933. The latent heat, in contrast to the Joule heat, is reconciled with reversibility since this heat is observed at the first-order phase transition which is reversible due to the equality of free energy (1).  

Hirsch knows thermodynamics better than others. Therefore the Hirsch Comment \cite{Hirsch2024Comment} once again indicates a regression in the understanding of thermodynamics. An open discussion of Hirsch's alternatives is important to overcome this regression, which began a hundred years ago because of unfounded belief in thermodynamics, in particular in the impossibility of violating the second law of thermodynamics. This open discussion may have both fundamental and practical importance, because of the importance of violating the second law of thermodynamics. The authors of a review article wrote: "{\it The second law of thermodynamics is, without a doubt, one of the most perfect laws in physics. Any reproducible violation of it, however small, would bring the discoverer great riches as well as a trip to Stockholm. The world's energy problems would be solved at one stroke}" \cite{PhysRep1999}. The world's energy problems can be solved since violation of the Carnot principle (7), which "{\it we call the second law of thermodynamics since Clausius's time}" \cite{Smoluchowski}, eliminates the need to use a fuel to create and maintain a temperature difference $T_{he} - T_{co}$ in a heat engine.

\end{document}